\documentclass[pre,aps,twocolumn,showpacs,preprintnumbers]{revtex4}

\usepackage{graphicx}
\usepackage{dcolumn}
\usepackage{bm}
\usepackage{amssymb}
\usepackage{pifont}
\usepackage{amsmath}
\usepackage{times}
\usepackage[nooneline]{subfigure}

\begin{document}
\title{Exponential localization of singular vectors in spatiotemporal chaos}

\author{Diego Paz\'o}
\affiliation{Instituto de F\'{\i}sica de Cantabria (IFCA), CSIC--Universidad de
Cantabria, E-39005 Santander, Spain}

\author{Juan M. L{\'o}pez}
\affiliation{Instituto de F\'{\i}sica de Cantabria (IFCA), CSIC--Universidad de
Cantabria, E-39005 Santander, Spain}

\author{Miguel A. Rodr{\'\i}guez}
\affiliation{Instituto de F\'{\i}sica de Cantabria (IFCA), CSIC--Universidad de
Cantabria, E-39005 Santander, Spain}

\date{appeared in Phys.~Rev.~E {\bf 79}, 036202 (2009)}

\begin{abstract}
In a dynamical system the singular vector (SV) indicates which
perturbation will exhibit maximal growth after a time interval
$\tau$. We show that in systems with spatiotemporal chaos
the SV exponentially localizes in space.
Under a suitable transformation, the SV  can be described in terms of
the Kardar-Parisi-Zhang equation with periodic noise. A scaling
argument allows us to deduce a universal power law $\tau^{-\gamma}$
for the localization of the SV. Moreover
the same exponent $\gamma$ characterizes the finite-$\tau$ deviation
of the Lyapunov exponent in excellent agreement with simulations.
Our results may help improving existing forecasting techniques.
\end{abstract}


\maketitle

\section{Introduction}
Lyapunov analysis is one of the main tools to
quantitatively characterize dynamical systems~\cite{eckmann}. In
particular, the largest Lyapunov exponent (LE) $\lambda$ is a
fundamental quantity that allows one to address, at some basic
level, the question of the predictability of a system. A positive
$\lambda$ implies chaotic behavior: neighboring initial conditions
diverge in time at an exponential rate $\sim \exp(\lambda t)$.
Actually, the LE is an asymptotic magnitude in the sense that it is
defined in the infinite-time limit.

However, it has been since long recognized that the infinite-time limit is often
irrelevant to address many important questions. Therefore, the {\em finite-time} LE
(to be defined below) was introduced as an essential tool to deal with problems
such as predictability~\cite{boffeta} and shadowability~\cite{sauer97} of
dynamical systems. Correspondingly, the singular vector (SV) (also referred to as the
finite-time Lyapunov vector) is defined as the perturbation that will 
show the maximal
expansion after some {\em finite} time $\tau$.

Singular vectors are typically used in operative models for weather
prediction~\cite{molteni96,Kalnay}. Every day the ECMWF (European Centre for
Medium-Range Weather Forecasts)~\cite{ecmwf} computes a set of
singular vectors to perform an ensemble forecasting of the European
weather~\cite{molteni96}. Also, in a different context, SVs have
been found to be useful to study advection of passive particles in
fluids~\cite{lapeyre02}.

It is a well-known result~\cite{goldhirsch87} that as the
the time horizon $\tau$ increases the SV exponentially
approaches the so-called forward Lyapunov vector
(FLV)~\cite{legras96}. For low dimensional systems
our understanding in these simple terms might be satisfactory. However, in
extended systems it is clear that spatio-temporal correlations
should have an important effect in the way this limit is achieved.
How this convergence of the SV toward the FLV occurs and what generic
system-independent features (if any) exist are important open
questions in the case of chaotic extended systems. Indeed, this is also
of practical and theoretical importance in the field of atmospheric
sciences~\cite{raro97,reynolds99}, because it is directly related
with the problem of realistic forecasting within a given finite time
horizon $\tau$.

In this paper we analyze the structure of the SV 
in extended systems with spatiotemporal chaos. We show that the SV localizes
exponentially in space around  
one center (determined by the present state of the system and by the
horizon $\tau$).
We find that the localization strength generically scales with the observation time
as a power-law $\tau^{-\gamma}$.
Moreover, we propose a stochastic
field-theoretical description of the evolution of the SV, which allows us to give an
analytical prediction for the universal exponent $\gamma$ and the finite-$\tau$
deviation of the LE, in good agreement with simulations in systems of different
natures, including coupled-map lattices and differential equations. Numerical results
together with general scaling arguments suggest our results are generic for a wide class
of systems exhibiting space-time chaos.

\section{Basic definitions}
For any given dynamical system,
infinitesimal perturbations ${\bm {\delta u}}(t)$ are governed by linear equations
(tangent space) such that the perturbation after some $\tau$ can be expressed via a linear
operator (resolvent or propagator): ${\bm {\delta u}}(t + \tau)=\mathrm{\mathbf{M}}(t +
\tau,t) {\bm{\delta u}}(t)$.

We are interested here in the SV, which is defined as the perturbation in tangent
space at time $t$ that gets
amplified the most at some future time $t+\tau$. In a more mathematical form, let
${\bm s}_\tau(t)$ denotes the SV for a time horizon $\tau$, then we have
\begin{equation}
{\bm s}_\tau(t)=\arg\max_{\bm {\delta u}(t)} \frac{\langle \mathrm{\mathbf{M}}(t+\tau,t)
{\bm {\delta u}}(t)\cdot
\mathrm{\mathbf{M}}(t+\tau,t) {\bm {\delta u}}(t) \rangle}
{\langle {\bm {\delta u}}(t) \cdot {\bm {\delta u}}(t) \rangle }
\label{eq_p}
\end{equation}
where $\langle{\bm x} \cdot {\bm y}\rangle$ denotes the scalar
product. The ``optimal'' perturbation ${\bm s }_\tau(t)$ is the SV
and it depends on both the time $t$ (that determines the
present state of the system), and the optimization time $\tau$.
Using the adjoint operator $\mathrm{\mathbf{M}}^*(t+\tau,t)$
defined by $\langle\mathrm{\mathbf{M}}(t+\tau,t) {\bm {\delta u}}(t)
\cdot \mathrm{\mathbf{M}}(t+\tau,t){\bm {\delta u}}(t) \rangle =
\langle {\bm {\delta u}}(t) \cdot \mathrm{\mathbf{M}}^*(t+\tau,t)
\mathrm{\mathbf{M}}(t+\tau,t) {\bm {\delta u}}(t) \rangle$, the
problem posed in Eq.~(\ref{eq_p}) can be solved by finding the
eigenvector of $\mathrm{\mathbf{M}}^*(t+\tau,t)
\mathrm{\mathbf{M}}(t+\tau,t)$ with the largest eigenvalue (all of
them are real and positive). For the Euclidean scalar product used
throughout this paper we have $\mathrm{\mathbf{M}}^* =
\mathrm{\mathbf{M}}^{\rm T}$.

The iterative application of the operator ${\mathbf \Phi}(t+\tau,t) \equiv
\mathrm{\mathbf{M}}^*(t+\tau,t) \mathrm{\mathbf{M}}(t+\tau,t)$ to an arbitrary vector (the
power method), generically converges to the wanted eigenvector ${\bm s}_\tau(t)$, which
satisfies the eigenvalue problem ${\mathbf \Phi}(t+\tau,t){\bm s}_\tau(t) = \mu_\tau(t)
{\bm s}_\tau(t)$. Then, the finite-time LE is typically defined as
$\lambda_\tau(t)=(2\tau)^{-1}\ln \mu_\tau(t)$~\cite{goldhirsch87,yoden93,okushima03},
which depends on $t$ and on the optimization time $\tau$.
According to Oseledec's theorem~\cite{oseledec}
in the limit $\tau
\to \infty$ with probability one $\lambda_\tau(t)$ converges to a time-independent
quantity-- namely, the largest LE of the chaotic attractor $\lim_{\tau \to
\infty} \lambda_\tau(t) = \lambda$.
An important observation is in order: $\lambda$ is usually obtained
from the evolution of a perturbation grown from the past, instead of considering
how a perturbation will evolve in the future.
Thus a perturbation evolved since the remote past is the 
``most unstable'' eigenvector of
$\mathrm{\mathbf{M}}(t,-\infty) \mathrm{\mathbf{M}}^*(t,-\infty)$ \cite{ershov98},
which has been called backward Lyapunov vector (BLV) \cite{legras96}. 
Conversely, the FLV is eigenvector of
$\mathrm{\mathbf{M}}^*(\infty,t) \mathrm{\mathbf{M}}(\infty,t)$.

In order to gain some numerical insight into the problem we have studied the behavior of
the SV in two prototypical dynamical systems exhibiting spatiotemporal chaos. The first
example we consider is a CML in a ring, which is discrete in
space ($x = 1, \ldots, L$) and time ($t = 0, 1, \ldots, \infty$):
\begin{eqnarray}
&u_x(t+1)= \epsilon f(u_{x+1}(t))+\epsilon
f(u_{x-1}(t)) + & \nonumber\\
& + (1 - 2\epsilon )f(u_x(t)),&
\label{cml}
\end{eqnarray}
with the logistic map $f(\varrho)=4\varrho (1-\varrho)$, and coupling
strength $\epsilon=1/3$.

The second model we consider here is an example of a chaotic system described by
differential equations. We study the model proposed by Lorenz in 1996 (L96)~\cite{lorenz96}
as a toy model in the context of weather dynamics. We consider the variables $y_x$ defined
in a ring geometry, and the evolution equations
\begin{equation}
\frac{d}{dt}y_x = -y_x-y_{x-1} (y_{x-2}-y_{x+1}) + F.
\label{l96}
\end{equation}
with $F=8$. We used the Euler method for the integration with $\Delta t= 10^{-4}$.

\section{Singular vector surface: localization, patterns and correlations}
An
important feature of singular and Lyapunov vectors is that their amplitudes are much localized
in space.
Localization of SVs was observed long time ago in simple meteorological
models~\cite{buizza93,yoden93}, however, to our knowledge, a description of the
localization profile in space or the dynamics does not exists even at a qualitative
level. In contrast, the localization properties of the BLV are much better understood.
Pikovsky {\em et al.}~\cite{pik94,pik98} demonstrated that, in a large class of
one-dimensional systems with spatiotemporal chaos, the BLV can be mapped into a
roughening surface that belongs to the universality class of the
Kardar-Parisi-Zhang (KPZ)
equation~\cite{kpz}. The solutions of the KPZ equation present universal scaling exponents
in space and time~\cite{kpz}. In particular, and as a result, one can conclude that the BLV
in one dimension is sub-exponentially localized, with its magnitude decaying in space as
$\sim\exp(-k \sqrt{x})$ from the localization site. 
These results apply to the FLV as well because the operator $\mathrm{\mathbf{M}}^*$
has the same coarse-grained long-scale structure than $\mathrm{\mathbf{M}}$,
and thus the FLV surface belongs to the KPZ universality class too.

The spatial structure of a Lyapunov vector is better resolved by making a logarithmic
transformation~\cite{pik94,pik98,sanchez04,szendro07}.
In the same spirit, expressing the vector components as
$\bm{s}_\tau(t) = [s_\tau(x,t)]_{x=1}^{x=L}$ we define a SV surface via the Hopf-Cole
transformation: $h_\tau(x,t)=\ln \vert s_\tau(x,t) \vert$.
In this representation the finite-$\tau$ LE becomes the average velocity of the surface:
$\lambda_\tau(t)=(\tau L)^{-1}\sum_{x=1}^{L}
\ln \vert [\mathrm{\mathbf{M}}(t+\tau,t) s_\tau(x,t)] / s_\tau(x,t) \vert$.
\begin{figure}
 \centerline{\includegraphics *[width=80mm]{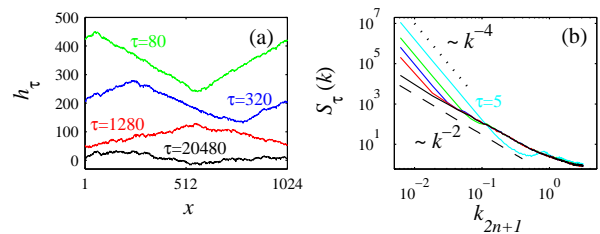}}
\caption{(Color online) (a) SV surfaces of the CML model for
different values of $\tau$ ($L=1024$). Curves have been shifted to
improve their visibility. (b) Structure factor for different values of $\tau$
(averaged over 500 realizations).\label{fig1}}
\end{figure}

Figure~\ref{fig1}(a) shows typical SV surfaces for different values
of $\tau$ in the case of the CML model (\ref{cml}). We obtained
similar behavior for the L96 model (\ref{l96}). We observe that for
$\tau$ larger than some threshold $\tau_\times(L)$ the corresponding
surface has approximately a random-walk profile; or in terms of the
SV itself we have stretched-exponential localization for a
a time horizon $\tau \gg \tau_\times(L)$, as anticipated
above.
Remarkably, for $\tau < \tau_\times(L)$ we find that the height
profile of the SV surface is facetted with superimposed fluctuations
at small scale. Note that a triangular SV surface corresponds to an
exponential localization of the SV.

The triangular pattern progressively dies out for longer times $\tau$.
This
means that the asymptotic convergence to the FLV
is reached for practical purposes in
a finite time: when we look for the perturbation at present time $t$ that will
exhibit the largest amplification at a future time $t+\tau$,
there is a characteristic time
$\tau_\times(L)$ that separates the genuine finite-time regime
from the quasi-infinite-time regime.
In the latter regime ($\tau > \tau_\times$) the SV has almost collapsed into
the FLV.

Quantitative information about the spatial correlations of the SV
can be obtained calculating the structure factor: $S_\tau(k)=
\lim_{t \to \infty} \langle\hat h_\tau(k,t) \hat h_\tau(-k,t)
\rangle$, where $\hat h_\tau(k,t)= L^{-1/2} \sum_x h_\tau(x,t)
\exp(ikx)$, and the bracket indicates average over realizations.
Figure \ref{fig1}(b) shows the results for different values of the
temporal horizon  $\tau$. For large $\tau$, one recovers the typical
spatial correlations corresponding to the FLV, $S_\tau(k) \sim
k^{-2}$, as expected from the previous discussion. Contrastingly, for $\tau <
\tau_\times(L)$ we find
$S_\tau(k) \sim k^{-4}$ at small wave numbers (due to the symmetry
of the triangular profile only wavenumbers $k_n=2\pi n/L$ with $n$
odd contribute significantly to the structure factor). The coarse-grained slope of the triangle $a$, and the
surface width $W^2_\tau(L) = L^{-1} \sum_x h_\tau(x,t)^2 - [L^{-1}
\sum_x h_\tau(x,t)]^2$ serve to quantify the localization strength.
The following relation holds for small $k$ and
$\tau < \tau_\times(L)$:
$$
S_\tau(k_n \to 0)\cong \frac{16 \langle a^2
\rangle}{L k_n^4} \cong \frac{768 \, \langle W^2_\tau(L)\rangle}{L^3
k_n^4}
$$
for $n$ odd.

\section{Stochastic field theory for singular vectors}
\begin{figure}
 \centerline{\includegraphics *[width=70mm]{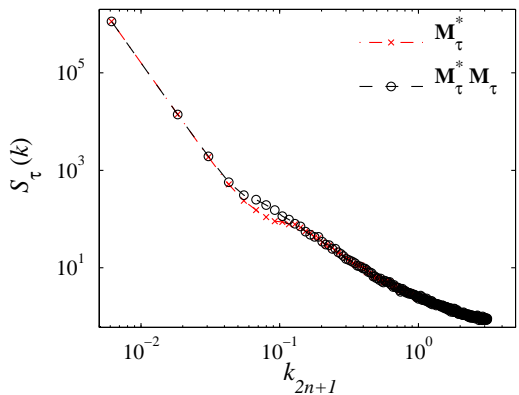}}
\caption{(Color online) Structure factor of the eigenvectors of
$\mathrm{\mathbf{M}}^*_\tau$ and
$\mathrm{\mathbf{M}}^*_\tau\mathrm{\mathbf{M}}_\tau$ for $\tau=160$
in the CML model after average over 500 realizations. The long scale
structure (small $k$) of the eigenvector of
$\mathrm{\mathbf{M}}^*_\tau\mathrm{\mathbf{M}}_\tau$ is well
captured by the eigenvector of $\mathrm{\mathbf{M}}^*_\tau$ alone.
\label{fig2}}
\end{figure}

SVs are the
asymptotic vectors computed by repetitive application of the
operator ${\mathbf \Phi}(t+\tau,t)=\mathrm{\mathbf{M}}^*(t+\tau,t)
\mathrm{\mathbf{M}}(t+\tau,t)$, to an arbitrary initial vector. This
makes the SV similar, albeit not exactly equivalent, to the BLV of a
periodic orbit. In fact, for temporally periodic but spatially
chaotic states the evolution in tangent space is mapped exactly
onto an one-dimensional localization problem and it can be
rigorously proved that all the eigenvectors are exponentially
localized~\cite{lepri96}. For the FLV to be computed, and because of
the $\tau\to \infty$ limit, only the $\mathrm{\mathbf{M}}^*$
operator needs to be used~\cite{ershov98}. However, the structure of
the operator ${\mathbf \Phi}$ is obviously different from that of
$\mathrm{\mathbf{M}}^*$. In Fig.~\ref{fig2} we compare the structure
factors obtained with the example of the CML when computing the
surfaces associated to the eigenvectors of ${\mathbf
\Phi}(t+\tau,t)$ (the actual singular vector) and
$\mathrm{\mathbf{M}}^*(t+\tau,t)$ (which can be viewed as the BLV of
an unstable periodic orbit of period $\tau$). One may see that, in
statistical terms, the large length scale behavior of the main
eigenvector of ${\mathbf \Phi}(t+\tau,t)$ is well reproduced using
operator $\mathrm{\mathbf{M}}^*(t+\tau,t)$ alone. This fact
justifies considering a minimal model (for the long scale structure
of the SV) consisting of a stochastic equation with {\em periodic}
noise.

We now write a minimal Langevin model to understand in the simplest terms
the structure of the SV surface on long
wavelengths. We propose a modification of the KPZ equation considering a periodic noise
(PNKPZ):
\begin{equation}
\partial_t h_\tau(x,t) = \zeta_\tau(x,t) + [\partial_x h_\tau(x,t)]^2 + \partial_{xx}
h_\tau(x,t) ,
\label{pnkpz}
\end{equation}
where one simply assumes $\zeta_\tau$ to be a random noise with
period $\tau$ [{\it i.e.} $\zeta_\tau(x,t) = \zeta_\tau(x,t+\tau)$]
and a $\delta$-correlator, $\langle\zeta_\tau(x,t) \,
\zeta_\tau(x',t') \rangle =2 \sigma \, \delta(x-x') \,
\delta(t-t')$, for $\vert t - t' \vert < \tau$.
\begin{figure}
 \centerline{\includegraphics *[width=80mm]{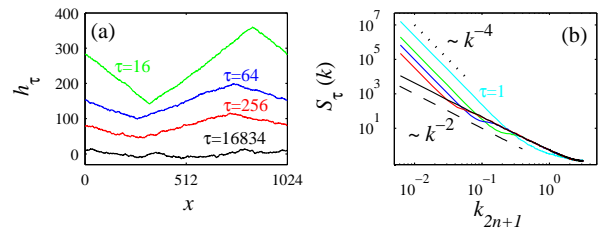}}
\caption{(Color online) (a) Asymptotic solutions of the PNKPZ equation for different values
of $\tau$ ($L=1024$). Curves are arbitrarily shifted to improve their visibility. (b)
Structure factors for different values of $\tau$ (averages were done over 500
realizations).\label{fig3}}
\end{figure}

Figure~\ref{fig3} shows the results of our numerical integration of Eq.~(\ref{pnkpz});
space and time were discretized ($\Delta x = 100\Delta t = 1$) and a noise
intensity
$\sigma=0.5$ was used. Other intensities can be used and this, of course, has no
effect on the results. One can see that the solutions of (\ref{pnkpz}) adopt a
triangular pattern, akin to those shown in Fig.~\ref{fig1} for the SV surface of the
deterministic model. The triangular structure flattens as $\tau$
increases and eventually disappears above some threshold value $\tau_\times(L)$.

\begin{figure}
 \centerline{\includegraphics*[width=80mm]{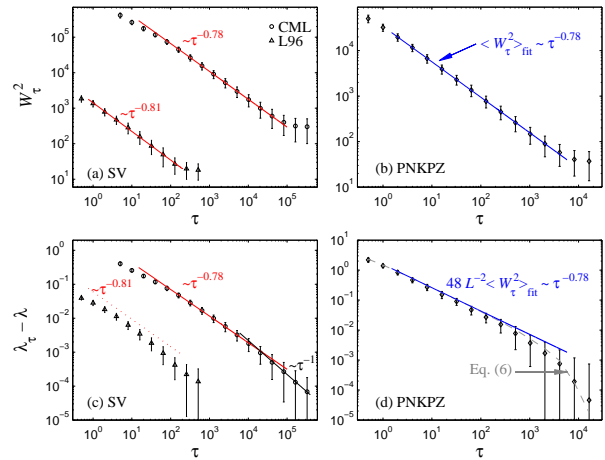}}
\caption{(Color online) (a,b) Surface width  vs.~$\tau$. (a) SVs of CML and L96 models for $L=4096$ and
$L=512$, respectively. (b) PNKPZ for $L=1024$. (c,d) Finite-$\tau$ deviation of
the LE. Data of the L96 model have been shifted two decades downwards to avoid
overlap of data sets. In all panels symbols mark mean values, and bars indicate the statistical
dispersion (100 realizations were done for the L96 model and 500 for the other models).
\label{fig4}}
\end{figure}

In Fig.~\ref{fig4}(a,b) we plot the surface width $W^2_\tau$ as
a function of $\tau$. Figure~\ref{fig4}(a) was generated from the SV surfaces of the
chaotic systems in Eqs.~(\ref{cml})-(\ref{l96}), while
Fig.~\ref{fig4}(b) was obtained from the numerical integration of the stochastic PNKPZ
equation (\ref{pnkpz}). For not too large values of $\tau$ we find a region of scaling
where the width is dominated by the triangular structure. We find 
$$
\langle W_\tau^2\rangle
\sim \tau^{-\gamma} 
$$
The values of $\tau$ with a $95\%$ of confidence are:
$\gamma=0.78\pm 0.01$ for the CML, $\gamma=0.81\pm 0.03$ for the L96 model, and
$\gamma=0.78\pm 0.01$ for the PNKPZ equation. This strongly suggests that $\gamma$ is a
universal exponent.

In all cases we expect the scaling to break down for large
$\tau$, as can be
inferred from the fact that the stochastic model (\ref{pnkpz})
converges to the KPZ problem as $\tau \to \infty$. This translates
into the convergence of the SV to the FLV as the time
horizon $\tau$ tends to infinity. The existence of the characteristic
time $\tau_\times(L)$ means that the FLV is virtually reached in a
finite time for a finite-size system.

For the theoretical analysis to follow, it is worth to remark that triangular
solutions also appear in the case of the quenched columnar KPZ (QCKPZ) equation discussed
in Ref.~\cite{szendro07b}, that corresponds to Eq.~(\ref{pnkpz}) but with the periodic
noise $\zeta(x,t)$ replaced with a quenched columnar disorder $\tilde\zeta(x)$.
Note that in the limit $\tau \to 0$ of the PNKPZ equation
[cf.~(\ref{pnkpz})] one recovers the QCKPZ equation. However, for a
finite $\tau$ correlations spread and one may conjecture that the
problem can be seen effectively as a QCKPZ equation where the disorder $\tilde\zeta(x)$
has a finite correlation length. The asymptotic solutions of the QCKPZ equation have been
described as a triangular pattern (with superimposed random fluctuations)
and exhibit scale-invariant roughness $\langle W_{\tau=0}^2(L) \rangle\sim L^{2\alpha_{\rm QC}}$~\cite{szendro07b}.

The existence of these two limiting dynamics ---namely, KPZ for
$\tau \to \infty$ and QCKPZ for $\tau \to 0$--- describes the
behavior of the PNKPZ equation and can be exploited to derive an
analytical expression for the exponent $\gamma$ by resorting to
scaling arguments as follows. On the one hand, the width of the
PNKPZ surface should scale as $\langle W_\tau^2(L)\rangle \sim
L^{2\alpha_{\rm KPZ}}$ for $\tau \gg \tau_\times(L)$, and it should
make a crossover to $\langle W_\tau^2(L)\rangle \sim L^{2\alpha_{\rm QC}}
\tau^{-\gamma}$ for $\tau \ll \tau_\times(L)$. Therefore, these two
limiting behaviors are separated by the crossover time
$\tau_\times(L) \sim L^{2(\alpha_{\rm QC}- \alpha_{\rm
KPZ})/\gamma}$. On the other hand, the characteristic time
$\tau_\times(L)$ should also correspond to the typical time that an
initially flat KPZ surface needs to reach the statistically
stationary regime in a system of size $L$, so we have
$\tau_\times(L)\sim L^{z_{\rm KPZ}}$ \cite{kpz}. Equating both
expressions we arrive at
\begin{equation}
\gamma=\frac{2(\alpha_{\rm QC}- \alpha_{\rm KPZ})}{z_{\rm KPZ}},
\label{gamma}
\end{equation}
which is in principle valid in {\em any} dimension and can be compared with our numerical simulations.
Inserting the critical exponents in one dimension ---$\alpha_{\rm KPZ}=1/2$ and $z_{\rm
KPZ}=3/2$~\cite{kpz}; and $\alpha_{\rm QC}=1.07 \pm 0.05$~\cite{szendro07b}--- we get
$\gamma=0.76\pm 0.07$ in good agreement with our simulations (our direct estimation of
$\alpha_{\rm QC}$ for the PNKPZ equation (\ref{pnkpz}) with $\tau=16$ yields $\alpha_{\rm
QC}=1.08\pm 0.02$, so $\gamma = 0.78\pm 0.03$).

\section{Finite-time Lyapunov Exponent}
A very interesting outcome of our analysis is the existence of a
scaling law for the finite-time LE. The average velocity of the
PNKPZ surface satisfies $\lambda_\tau(t)=\langle
L^{-1}\sum_x{(\partial_x h_\tau(x,t))^2} \rangle_\tau$, where in
good approximation the noise term has been assumed to average out
within a period. Existing results on QCKPZ~\cite{szendro07b}
indicate that there is a complete separation in two independent
components such that $h$ can be constructed as the sum of a
triangular pattern with a tilt $\pm ax$ and the solutions in $\tau
\to \infty$. After some algebra, one obtains
\begin{equation}
\langle \lambda_\tau \rangle -\lambda =  \langle a^2 \rangle
= 48 L^{-2}(\langle W_\tau^2 \rangle - \langle W_{\tau=\infty}^2\rangle)
\label{finite_le}
\end{equation}
In Fig.~\ref{fig4}(d) we test the validity of this expression by
plotting $\lambda_\tau -\lambda$ as a function of $\tau$. The dashed
line arises inserting in (\ref{finite_le}) the mean values of
$W_\tau^2$ depicted in Fig.~\ref{fig4}(b). The straight line is
computed by inserting in the dominant contribution $48 L^{-2}
\langle W_\tau^2\rangle$ the fitting power law in
Fig.~\ref{fig4}(b): $\left< W_\tau^2 \right>_{\rm fit} = 3.4 \times
10^{4} \, \tau^{-0.78}$. The agreement is, in our opinion, very good
taking into account the simplicity of our arguments.

For a generic system with spatiotemporal chaos we may not expect to be able to obtain a
formula relating the width of the SV surface and the finite-$\tau$ deviation of the LE,
but still we may ask whether the $\tau^{-\gamma}$ dependence is observed.
Figure~\ref{fig4}(c) indicates that this scaling for the LE exists in the same
region where $\langle W_\tau^2\rangle \sim \tau^{-\gamma}$, although the result is not
conclusive for the L96 model because computer limitations
did not allow us to study larger systems.
Finally, we point out that for  $\tau > \tau_\times(L)$ the convergence of $\lambda_\tau$
to $\lambda$ becomes faster, precisely as $\sim\tau^{-1}$ in accordance with the standard
asymptotic behavior for low-dimensional systems~\cite{goldhirsch87}. This means that
the spatial degrees of
freedom slow down the convergence to the LE in extended dynamical systems
as compared with the low-dimensional case.

\section{Conclusions}
We have shown that for a large family of systems the SV exhibits
exponential localization.
We claim that this spatial structure can be described in terms of the
solutions of the a stochastic equation, related to KPZ
equation, in which the noise is periodic or quenched. In terms of the
associated surfaces, the SV converges toward the FLV following a
power law with a universal exponent $-\gamma$ given by
Eq.~(\ref{gamma}). In particular, for one-dimensional systems one
has $\gamma\simeq 0.78$. Moreover, the same exponent characterizes
the finite-time deviation of the LE. Even for spatio-temporal
chaotic systems whose BLV (and FLV) does not belong to the universality class
of KPZ, as occurs with Hamiltonian lattices~\cite{pik01},
one may expect a qualitative similarity with the scenario presented in this 
work, but possibly with different exponents.

In forecasting applications $\tau$ is the control parameter to
calibrate the ensemble of SVs. Our results show how $\tau$
quantifies (through the universal exponent $\gamma$)
the localization strength, and the exponential growth rate.

\acknowledgments
Financial support from the Ministerio de Educaci\'on y Ciencia (Spain) under projects
FIS2006-12253-C06-04 and CGL2007-64387/CLI is acknowledged.
D.P.~acknowledges the support by CSIC under the Junta de Ampliaci\'on de
Estudios Programme (JAE-Doc).

\end{document}